# A Historical Perspective on the Topology and Physics of Hyperspace

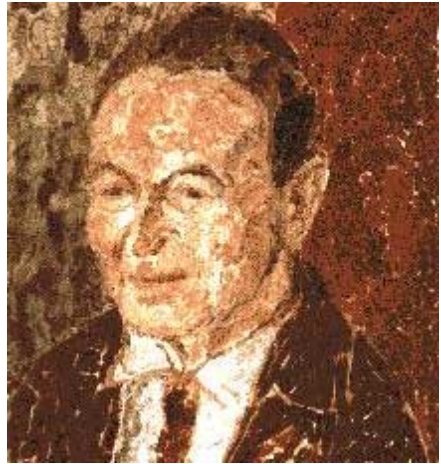

**Ian T. Durham**

> **di•men•sion** *n* **1 a ...** (2) : one of a group of properties whose number is necessary and sufficient to determine uniquely each element of a system of usu. mathematical entities (as an aggregate of points in real or abstract space) <the surface of a sphere has two ~s>; *also* : a parameter or coordinate variable assigned to such a property <the three ~s of momentum> **...** [1]

## Introduction

Throughout history, the human mind has sought to understand its surroundings. One of the most fundamental aspects of our universal surroundings is the array of spatial and temporal dimensions within which we exist. Humanity has slowly and discontinuously managed to unfold eleven (or twelve) of these dimensions over the last 2550 years or so. In this paper the historical development of the mathematics and physics behind the discovery of these dimensions is examined from the earliest records of Greek geometers and scientists starting in the sixth century BCE through to the most recent developments in theoretical physics. Historical glimpses of the people who have helped to shape these developments are given as a basis for the mathematical processes that build up to the overall worldview. It is wise to note that there were many parallel developments, notably during ancient times, in the Middle East and Far East. For the sake of brevity they have not been included here, however, the reader is encouraged to explore these areas in greater depth. In particular, the legacy of Euclid and Klein are developed in depth working from the Euclidean concept that the topology of the universe is inherently flat and moving into Klein's first use of a curved dimension in modifying Kaluza's initial work on five dimensions. This paper relies heavily on secondary sources as it is merely meant to be an introduction to the topic.

## Early Developments

The earliest developments in Greek mathematics are attributed to Pythagoras and his followers. Pythagoras was born around 570 BCE on the island of Samos off the Ionian coast. He supposedly left the island around 540 BCE out of disenchantment with the ruling Polycrates and fled to Croton. Croton was a Greek settlement on the southeastern coast of Italy on the lower Adriatic. Once in Croton he attracted a group of followers who have since been known as the Pythagoreans, a mysterious group who have been revered and copied, reportedly, by druids, masons, and secret societies over the centuries. Their teachings and, as a result, those of Pythagoras himself were kept secret. Most knowledge of their teachings was not revealed until nearly a century later in the writings of Philolaus. Thus the teachings are far from a direct account of Pythagorean thought. Recent tradition even indicates Pythagoras may have learned much of his teachings from other sources, possibly on a series of travels he had undertaken. It is said these travels were in the East, though there is a persistent legend that quotes him as saying:

---

[1] F.C. Mish Ed., *Webster's Ninth New Collegiate Dictionary*, Merriam-Webster, 1991.

> All I know I learned from a Druid.[2]

The earliest written evidence of Greek mathematics is in fact Euclid's *Elements*. This work dates from the fourth century BCE though much of it is thought to be the work of earlier mathematicians, including the Pythagoreans. Eudemus the Peripatetic attributes to them the theorem that describes the sum of the interior angles in a triangle being equal to the sum of two right angles. His description of the Pythagoreans' proof of the theorem is as follows:

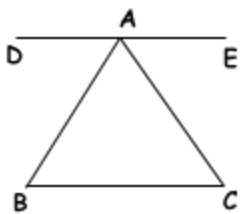

> Let *ABC* be a triangle, and let the line *DE* be drawn through *A* parallel to *BC*. Now since *BC* and *DE* are parallel, and the alternate angles are equal, the angle *DAB* is equal to the angle *ABC* and the angle *EAC* is equal to the angle *ACB*.
>
> Let the angle *BAC* be added to them both. Then the angles *DAB, BAC,* and *CAE* (that is to say, the angles *DAB* and *BAE*, i.e., two right angles) are equal to the three angles of the triangle *ABC*.
>
> Hence the three angles of the triangle are equal to two right angles.[3]

According to this proof, the Pythagoreans were already familiar with the concept of parallel lines as well as two-dimensional objects. Certainly, humans were aware of two dimensions from simple sensory perceptions, however, this proof is one of the earliest pieces of evidence that indicates an understanding of the mathematical nuances lying behind the physical reality.

Of course, the more famous mathematical construct attributed to the Pythagoreans is the famous Pythagorean theorem that describes the sum of the squares on the shortest two sides of a right triangle as being equal to the square of the hypotenuse, or longest side. The Pythagoreans' use of one-dimensional lines laid down to describe a two-dimensional feature was one of the earliest examples of extending a single dimension in a way that creates an extra or "higher" orthogonal dimension. In fact, if a single straight line is considered to be one-dimensional, and a second straight line, non-parallel to the first, is laid down and connected at any point to the first, a two-dimensional space is *automatically* created where a minimum of two position coordinates must be specified in order to describe a single point in that space. Realizing that such abstract spaces as Riemannian geometry had yet to be developed, this can be considered one of the first mathematical realizations of multi-dimensional space.

---

[2] A legendary quote often found in popular accounts and products relating to the druids and druidry.
[3] Proclus in *Euclid I*, qtd. in J.M. Robinson, *An Introduction to Early Greek Philosophy*, Houghton Mifflin, 1968.

In contrast, other Greek geometers used a method that consisted of applying a method of "area mathematics" to decipher algebraic problems. Since they did not possess a form of algebra similar to our own, this method of area application allowed them to solve second-degree equations and formed the basis of Euclid's work on irrationals.

## Euclid's *Elements*

Euclid was born around 325 BC. Not much is known of his life other than the fact that he lived and worked in Alexandria, Egypt. From various accounts including those of Proclus, Euclid compiled and refined the work of many of his predecessors in his famous anthology, *The Elements*. The exact nature of his personal contribution and ideas is sketchy. However, Proclus wrote that Euclid also brought "to irrefutable demonstration the things which had been only loosely proved by his predecessors."[4]

Euclid's *Elements* begins by defining certain terms vital to the understanding of geometrical space. It is important to make note of a few of these definitions in our study of advancing dimensions.

First, Euclid defines a point as "that which has no part."[5] He also defines a line as a "breadthless point"[6] asserting the Pythagorean use of lines as one dimensional objects. In addition, the definition of a point is of particular importance as it is one of the earliest assertions that zero-dimensional objects can be represented mathematically. As we will see, this plays an important role later in the definition of string theory as the original concept of point-particles based on the Euclidean assumption of an indivisible point is amended and the actual definition of "point-like" is no longer quite the same thing.

Euclid defines a surface as "that which has length and breadth only."[7] It is interesting to note that Euclid differentiated between this simple definition of a surface and that of a plane surface which he defined as "a surface which lies evenly with the straight lines on itself."[8] What is interesting to note is that the definition of height (as a separate concept from length and breadth) is not given until Book VI and is given completely independently of length and breadth as "the perpendicular drawn from the vertex to the base."[9]

---

[4] Euclid in *Elements I*, D. E. Joyce Ed., Clark University Mathematics Department Web, 1998.
[5] Ibid.
[6] Ibid.
[7] Ibid.
[8] Ibid.
[9] Euclid in *Elements VI*, D. E. Joyce Ed.

Euclid's first indication that a *third* spatial dimension exists specifically and mathematically in relation to the first two spatial dimensions, comes in Book XI. However in Book VII he states that "when three numbers having multiplied one another make some number, the number so produced be called *solid*, and its *sides* are the numbers which have multiplied one another."[10] The definition in Book XI is the more familiar one. Here a solid is defined as "that which has length, breadth, and depth."[11] Further, a surface is defined as the side of a solid. Thus Euclid took the Pythagorean concept a step further (at least in 'print') and created a third dimension orthogonal to both the previous two. Relating this to my assertion that by simply creating two non-parallel, yet connected lines, a two-dimensional space is immediately created, to ensure the existence of the third dimension in an additional space, three lines must be connected (only once each) and be completely non-parallel, and one must be non-coplanar, to each other to ensure three spatial dimensions. The easiest way to do both of these extensions is to make the lines *themselves* completely orthogonal, thus creating a visual aide in perceiving three orthogonal dimensions much as we create $x$, $y$, and $z$ axes when plotting a point in three-dimensions (it should be noted, the dimensions are always orthogonal, but to create them, the lines only need to be non-parallel, with one non-coplanar, and connected once – a moment or two of thought should confirm this).

It is interesting to note that nowhere does Euclid define addition and subtraction. He assumes that these basic functions are known and understood. However, multiplication is specifically defined. Euclid also represents numbers solely in the context of a line while it was apparent that the Pythagoreans represented numbers as figures.[12]

Thus, by the end of the third century BCE, the three spatial dimensions as we can perceive them, were mathematically known and rigorously defined. Euclidean geometry then remained the only accepted description of the spatial universe until well into the Renaissance in the 17$^{th}$ century CE.

There is one aspect of Euclid's work that bedeviled mathematicians and physicists for nearly two millennia. This is frequently referred to as Euclid's *fifth postulate*. The fifth postulate states "that if a straight line falling on two straight lines makes the interior angles on the same side less than two right angles, the two straight lines, if produced indefinitely, meet on that side on which the angles are less than the two right angles."[13] This, it turns out, is the essence of Euclidean geometry. It assumes a *non-curved* space in all dimensions. Every other postulate of Euclidean geometry can hold true on certain surfaces except the fifth postulate. Mathematicians struggled to prove or disprove or disassociate this principle until well into the 19$^{th}$ century when it seems Gauss was the first to accept the possibility of non-Euclidean geometry and, in conjunction with Bolyai and Lobachevski, showed through the development of curved geometrical spaces, that

---

[10] Euclid in *Elements VII*, D. E. Joyce Ed.
[11] Euclid in *Elements XI*, D. E. Joyce Ed.
[12] D. E. Joyce, *Guide to Euclid's Elements VII*, Clark University Mathematics Department Web, 1998.
[13] Euclid in *Elements I*, D. E. Joyce Ed.

Euclid's fifth postulate was indeed independent.[14] This opened the door to a vastly new way of representing dimensions. We will see that this has an important impact on the discovery of the dimensions beyond the fourth (time) and third (spatial). The difficulty in proving this postulate also secured Euclid's legacy for two millennia as dimensions were seen as completely flat. It should be noted that Farkas Bolyai, a geometer in his own right and a lifelong friend of Gauss, attempted, in vain, to stop his son Janos from contemplating this problem, but, luckily, failed in his attempt.[15]

## A Brief Foray into Phase Space

The dimensions described here are limited to actual physical dimensions. However, the use of a mathematical tool called phase space has utilized the concept of a dimension as a direction orthogonal to other defined directions in a unique and handy way. Phase space is defined as the number of dimensions that can be utilized to represent the state of a particular system at a given time and is equal to the number of degrees of freedom of the that system. Frequently momentum is the quantity represented in addition to our customary space and time, although, in the first known representation of its kind, in 1698 by Varignon, velocity was instead used in place of momentum[16]. The important distinction of phase space is that it is a useful mathematical tool but does not describe dimensions in the same way used here. It merely allows for an easier representation of the state of a particular system at a given time. This is particularly useful in the representation of quantum states where momentum (through the uncertainty principle) plays an important role in the description of the given state. This actually does end up playing an important role later on as we will see that string theory relies heavily on the principles outlined by the uncertainty principle and draws on the fundamentals of quantum mechanics. However, strictly speaking, it is not the type of dimension we are interested in. We will pay close attention only to those dimensions that are physical actualities and allow the physical transfer of energy in one or two directions within that dimension (finding a clear cut definition of a dimension in physics is not an easy task, as we will see, particularly in the context of temporal dimensions) and that can be represented in length units alone (this includes temporal dimensions).

## Time as a Dimension

The use of time as a dimension in mathematical plots dates from well before the 19th century. It was in the latter portion of this century, however, when the use of time as a true dimension, able to be represented by a *length*, and consistent with the previous uses of spatial dimensions, was brought to bear. Actually, for the first representation of time as an independent orthogonal coordinate in a four-dimensional space-time, we must look to the early 20th century. In fact, Einstein used a purely algebraic form of math to describe special relativity in 1905. It was not until 1908 that time was included as a coordinate in a four-dimensional space-time by Minkowski.[17] This gave rise to the geometrical representation of relativity and led directly to the development of general relativity in 1914-16. It is imperative to remember, however, that until we reach Klein's formation of the fifth dimension, we are still in Euclidean space-time which means all dimensions are still flat.

---

[14] S. Weinberg, *Gravitation and Cosmology: Principles and Applications of the General Theory of Relativity*, John Wiley & Sons, 1972.
[15] http://www-history.mcs.st-andrews.ac.uk/history/Mathematicians/Bolyai_Farkas.html
[16] J. Stachel, *Einstein: A Man for the Millennium?*, lecture to the Spring 2000 New England Section meeting of the American Physical Society and American Association of Physics Teachers.

At first, it is not imperative to represent time in spatial units to visualize the logical extension of space to space-time. However, in order to do any meaningful mathematics, it would be desirable to develop some way in which to represent time with the same unit measurement as space. Building on Einstein's postulate that the speed of light is invariant and universal, it can be represented as a unitless number. The most logical choice of number in this case is 1. Therefore, if the speed of light is $c$ then $c = 1$.

More specifically, the speed of light is defined as:

$$c = \frac{\text{distance light travels in any given time interval}}{\text{the given time interval}}$$

In order to ensure that this number is unitless, we must represent time spatially. So, in SI units, applied to relativity, time is measured in meters. This representation is often referred to as *natural units*. This allows us to construct yet another useful tool in relativity: the space-time diagram. In this diagram, a spatial coordinate (usually x in two dimensions) is plotted as one axis with time as the other, both being represented in natural units as meters:

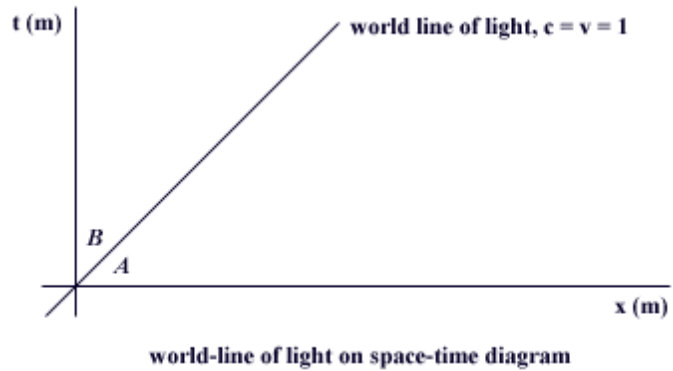

**world-line of light on space-time diagram**

In this diagram, the angles *A* and *B* are both equal to 45°. This equality represents the invariability of the speed of light. The time and space axes are not always at right angles in this diagram but they cannot 'pass through' the world-line of light which represents a barrier. The slope of this line is $dt/dx = 1/v$. Given an event (any single point on this diagram), the interval between any two events separated by coordinate increments (Δt, Δx, Δy, Δz) is given as:

$$\Delta s^2 = -(\Delta t)^2 + (\Delta x)^2 + (\Delta y)^2 + (\Delta z)^2 \qquad (1).$$

---

[17] B. F. Schutz, *A First Course in General Relativity*, Cambridge University Press, 1990.

This interval is invariant and forms the basis for building a space-time metric. The Minkowski metric which we use in the still flat space-time of special relativity is defined as:

$$\eta_{\alpha\beta} = \begin{pmatrix} -1 & 0 & 0 & 0 \\ 0 & 1 & 0 & 0 \\ 0 & 0 & 1 & 0 \\ 0 & 0 & 0 & 1 \end{pmatrix} \qquad (2).$$

So in brief review, by 1908, we have seen the recognition of four dimensions beginning with Pythagorean representations of two-dimensional objects and properties, working through Euclid's definitions surrounding three-dimensional objects, and finally reaching, over two millennia later, the representation by Minkowski and Einstein of time as the fourth dimension.

It is presumably safe to say that the four dimensions of space and time as presented here, represent the limit of human sensory perception. This limitation is most likely the reason that over two thousand years passed between the establishment of the initial three spatial dimensions and the addition of the next spatial dimension. The perception of time as a dimension could quite possibly be viewed as a concept behind its time. It certainly contains a sophisticated level of mathematics, but is consistent enough in its linearity that it could be considered a mere fluke that it had not been perceived as a dimension by even the Greeks. Evidence points to its having appeared on the same plot as spatial coordinates two-hundred years prior to Minkowski's use of time as a coordinate. However, ignoring this fact, the limit of human sensory perception can be considered the greatest barrier that needed to be overcome in order to even conceptualize higher dimensions. Einstein's relativity opened the door for this, a lifelong dream of Riemann, but it was an obscure mathematician named Theodor Kaluza who first mathematically developed the idea (Nordstrom was also successful in this effort, but has not been the beneficiary of a 'named' theory in this area).

**The Fifth Dimension**

We would be remiss, however, if we do not mention the fact that two rather colourful scientists publicly proposed the fifth dimension (they referred to it as the fourth dimension as Minkowski's work had yet to be published) less than thirty years prior.

In 1877, a bizarre and sensational trial took place in London. The then renowned psychic Henry Slade sat accused of fraud for supposedly deceiving his clients who were some of England's elite. Quite possibly the most bizarre part of the trial was the fact that several prominent physicists of the time, including some future Nobel Prize winners, came to Slade's defense by supporting the notion of a 'fourth dimension' (spatially speaking – for our sake, it is the fifth dimension). One of these physicists was Johann Zollner a professor of physics and astronomy at the University of Leipzig. Zollner enlisted the aide of famous physicists William Crookes, Wilhelm Weber, J.J. Thompson, Lord Rayleigh, and others in an attempt to prove Slade's innocence. Slade was ultimately convicted, at no surprise to us (for we now know the amount of energy required to manipulate the 'fourth dimension'), but Zollner was so convinced of the existence of the 'fourth dimension' and its ability to be manipulated that he published articles in both scientific and pseudo-scientific journals in defense of it.[18]

In the very same year as the Slade trial, a mathematician named Charles Howard Hinton graduated from Oxford. Being the son of famous ear surgeon and renowned bigamist James Hinton, his personal life began as anything but dull. He eventually became a bigamist himself, having taken the widow of George Boole (of Boolean algebra fame) as his first wife, and Maude Weldon as his second. Despite his arrest, his first wife, Mary, declined to press charges and they both fled to the United States. It was here that Hinton eventually found his way from Princeton to the US Naval Observatory and finally, to the place where another great physicist of the time was 'born:' the patent office (though not the same patent office, of course). Hinton was known as the man who could 'see' the fourth dimension and spent his life laboring to develop ingenious visual descriptions of the fourth dimension. These descriptions eventually became known as hypercubes and unraveled hypercubes became known as tesseracts, a term coined by Hinton himself.[19]

But, despite all this laboring on the part of physicists and mathematicians during the late 19th century, it was not until after Einstein published his seminal work on general relativity that Theodor Kaluza was able to become to the first to mathematically describe the fifth (or fourth spatial) dimension.

Theodor Kaluza was born in 1885 in Ratibor, Germany, now known as Raciborz, Poland, eight years after the famous Slade trial. He was a professor at Königsberg when in 1919 he sent Einstein a paper he had been working on that unified Einstein's relativity with Maxwell's theory of light. The very means for unifying these two theories was the addition of a fifth dimension. What separated Kaluza's work from that of Riemann, Zollner, and Hinton, was that Kaluza was proposing a true field theory. He simply wrote down Einstein's field equations in five dimensions. He then showed that the new five-dimensional equations contained Einstein's four-dimensional relativity theory plus an additional piece. It turned out the additional piece was exactly Maxwell's theory of light. In Kaluza's original theory, all the fields involved were independent of the fifth dimension. By starting with pure gravity written in five dimensions, though independent of the fifth dimension, the field breaks down to four dimensions which ultimately leaves a metric, a Maxwell field, and a scalar. Kaluza's major restriction on the fifth dimension was that it was cylindrical in form thus forcing it not to appear in the physics of the problem (i.e. it was a convenient mathematical device, but held little real meaning).

---

[18] M. Kaku, *Hyperspace: A Scientific Odyssey Through Parallel Universes, Time Warps, and the 10th Dimension*, Anchor Books, 1994.
[19] Ibid.

Oskar Klein (born in 1894), then professor at the University of Michigan, refined Kaluza's ideas in 1926. These combined works are now known as Kaluza-Klein theory. Klein did not assume total independence of the fifth dimension. Returning to Euclid's fifth postulate, we recall that Gauss, Bolyai, and Lobachevski were the first to prove the independence of Euclid's fifth postulate. This opened the door to curved geometries. Kaluza had initially employed a cylindrical shape (topology) for his fifth dimension, but it was still Euclidean in geometry. Klein utilized the new idea of non-Euclidean geometry and postulated that Kaluza's fifth dimension was actually curved in geometry and microscopic in size. In fact, Klein assumed this dimension would have the topology of a circle with the radius on the order of the Planck length (we will see the exact radius becomes important in string theory). We can then write the topology for all five dimensions as $B^4 \times S^1$ where the fifth coordinate, y, is periodic – $0 \leq my \leq 2\pi$ - and m is the inverse radius of the circle. The periodicity of the extra dimension allows us to make a Fourier expansion in this coordinate. The first order terms of this expansion correspond to the reduction initially introduced by Kaluza.

Working with the convention adopted by Derix and van der Schaar[20] we will define hatted quantities as being five-dimensional and unhatted quantities as four-dimensional. Five dimensional indices will run as: $\hat{\mu} = 0,1,2,3,5$ and the four-dimensional indices will run as: $\mu = 0,1,2,3$ ( $x^{\hat{\mu}} = (x^{\mu}, y)$ ).

Kaluza wrote the five-dimensional metric as follows, with a 4+1 split:

$$\hat{g}_{\hat{\mu}\hat{v}} = \begin{pmatrix} g_{\mu v} - \phi A_\mu A_v & -\phi A_\mu \\ -\sigma A_v & -\phi \end{pmatrix} \qquad (3)$$

This allows the four-dimensional fields to have the proper transformation characteristics in four dimensions. As developed by Derix and van der Schaar, we must first consider an infinitesimal coordinate transformation in five dimensions:

$$x^{\hat{\mu}} \to x^{\hat{\mu}} + \varepsilon \xi^{\hat{\mu}}(x^\mu)$$

where the transformation is independent of the fifth coordinate. Given this coordinate transformation, we can transform the five dimensional metric in the following way:

$$\partial \hat{g}_{\hat{\mu}\hat{v}} = \hat{g}_{\hat{\mu}\hat{\rho}}\left(\partial_{\hat{v}}\xi^{\hat{\rho}}\right) + \hat{g}_{\hat{\rho}\hat{v}}\left(\partial_{\hat{\mu}}\xi^{\hat{\rho}}\right) + \xi^{\hat{\rho}}\left(\partial_{\hat{\rho}}\hat{g}_{\hat{\mu}\hat{v}}\right) \qquad (4).$$

---

[20] M. Derix and J. P. van der Schaar, *Stringy Black Holes*, Master's Thesis, University of Groningen, 1998.

We can then derive the transformation properties of the four-dimensional vector $A_\mu$ as follows:

$$\partial \hat{g}_{\mu 5} = -(\partial \phi) A_\mu - \phi (\partial A_\mu)$$
$$= \hat{g}_{\hat{\rho} 5}(\partial_\mu \xi^{\hat{\rho}}) + \xi^{\hat{\rho}}(\partial_{\hat{\rho}} \hat{g}_{\mu 5})$$
$$= -\phi A_\rho (\partial_\mu \xi^\rho) - \phi(\partial_\mu \xi^5) - \xi^\rho (\partial_\rho \phi) D_\mu - \xi^{\hat{\rho}}(\partial_{\hat{\rho}} D_\mu)\phi$$

therefore

$$\partial A_\mu = A_\rho (\partial_\mu \xi^\rho) + \xi^\rho (\partial_\rho A_\mu) + \partial_\mu \xi^5 \qquad (3).$$

The last term in this equation is a $U(1)$ gauge term and $A_\mu$ has right transformation properties in four dimensions. The invariance of general coordinates in five dimensions and the independence of the fifth dimension (still held by Klein despite his topology change from flat to curved) results in gauge symmetry of the four-dimensional vector. The gauge symmetries become more complicated in four dimensions and are a result of more complicated compactifications, which is an important part of string theory and Calabi-Yau spaces (as we will later see).

The four-dimensional metric and scalar also have the correct transformations:

$$\partial g_{\mu\nu} = g_{\mu\rho}(\partial_\nu \xi^\rho) + g_{\rho\nu}(\partial_\mu \xi^\rho) + \xi^\rho (\partial_\rho g_{\mu\nu})$$

and

$$\partial \phi = \xi^\rho \partial_\rho \phi.$$

Here, Derix and van der Schaar have set $\hat{g}_{55} = -\phi$ which keeps the scalar field positive while also keeping the fifth coordinate space-like. Keeping in mind the fact that $\hat{g}_{\hat{\mu}\hat{\rho}} \hat{g}^{\hat{\rho}\hat{\nu}} = \partial_{\hat{\mu}}^{\hat{\nu}}$ the inverse metric can be written as:

$$\hat{g}^{\hat{\mu}\hat{\nu}} = \begin{pmatrix} g^{\mu\nu} & -A^\mu \\ -A^\nu & -\frac{1}{\phi} + A^2 \end{pmatrix} \qquad (4).$$

To develop Kaluza's idea, we begin with pure gravity, meaning a source-free space-time in five dimensions. The action integral for this system is given by Derix and van der Schaar[21] as:

$$S^{(5)} = -\int d^5 x \sqrt{\hat{g}} \hat{R} \qquad (5).$$

The constant in front of the integrals in equation 6 can be inserted here, but was left out by Derix and van der Schaar. See Overduin and Wesson for a more in depth discussion of this.[22] Compare this to the action integral given by Visser[23] (for comparison, see those given by Weinberg[24] and Misner, Thorne, and Wheeler[25]) from which we can derive general relativity in four dimensions:

$$S = -\frac{c^3}{16\pi G} \int_\Omega R\sqrt{g} d^4 x - \frac{c^3}{8\pi G} \int_{\partial\Omega} K\sqrt{^3g} d^3 x + \int_\Omega L\sqrt{g} d^4 x \qquad (6).$$

The four dimensional action integral is far more complicated than the five dimensional one. This is one of the most important aspects of multi-dimensional physics. In this way, physicists have used the addition of extra dimensions to simplify complex mathematical problems, the most important example being string theory.

Returning to the five dimensional model, the determinant of the metric can be reduced to:

$$\hat{g} = \det(\hat{g}_{\hat{\mu}\hat{\nu}}) = -\det(g_{\mu\nu})\phi = -g\phi \qquad (7).$$

Derix and van der Schaar present the result of the derivation of the Ricci curvature scalar in five dimensions as:

$$\hat{R} = R + \frac{1}{2\phi^2}(\partial\phi)^2 - \frac{1}{\phi}\Box\phi + \frac{1}{4}\phi F_{\mu\nu}(A)F^{\mu\nu}(A) \qquad (8)$$

where $F_{\mu\nu} = \partial_\mu A_\nu - \partial_\nu A_\mu$. Putting this back into equation 5 and assuming that integration over the fifth coordinate is 1 ($dx^5 = 1$), the action becomes:

$$S^{(4)} = \int d^4 x \sqrt{-g\phi} \left\{ -R - \frac{1}{2\phi^2}(\partial\phi)^2 + \frac{1}{\phi}\Box\phi - \frac{1}{4}\phi F(A)^2 \right\} \qquad (9).$$

Both terms involving derivatives of $\phi$ can be written as total derivatives thus not contributing to the action and simplifying equation 9 to:

---

[21] Ibid.
[22] J. M. Overduin and P. S. Wesson, *Kaluza-Klein Gravity,* Phys. Rep. 283, 303, 1997.
[23] M. Visser, *Lorentzian Wormholes: From Einstein to Hawking*, AIP Press/Springer, 1996.
[24] S. Weinberg, 1972.
[25] C.W. Misner, K.S. Thorne, and J.A. Wheeler, *Gravitation*, W.H. Freeman & Company, 1973.

$$S^{(4)} = \int d^4x \sqrt{-g}\, \phi^{1/2} \left\{ -R - \frac{1}{4}\phi F(A)^2 \right\} \quad (10).$$

Derix and van der Schaar take this a few steps further by deriving a form involving an Einstein term and additional exotic matter terms. They first do this by performing a conformal rescaling of the metric:

$$g_{\mu\nu} \to g'_{\mu\nu} = \phi^{\frac{1}{2}} g_{\mu\nu}$$

Non-trivially, the Ricci scalar transforms to the following four-dimensional form:

$$R = \phi^{\frac{1}{2}} \left[ R' + \frac{3}{2}\left( \nabla_\rho \left(\frac{1}{\phi}\nabla^\rho \phi\right) - \frac{1}{4\phi^2}(\nabla\phi)^2 \right) \right] \quad (11).$$

Finally, they transformed the other terms in the action as follows:

$$F^2 = \phi F'^2,$$

$$\sqrt{-g} = \phi^{-1}\sqrt{-g'},$$

$$\phi \to \phi' = \sqrt{3}\log\phi.$$

Finally, the four-dimensional action can be written in the conventional form (Derix and van der Schaar dropped the primes):

$$S = \int d^4x \sqrt{-g} \left\{ -R + \frac{1}{2}\partial_\mu \phi \partial^\mu \phi - \frac{1}{4}e^{-\sqrt{3}\phi} F_{\mu\nu} F^{\mu\nu} \right\} \quad (12).$$

A detailed construction of the action for gravitational fields is contained in Chapter 12 of Weinberg.[26] In this way, we see that by adding a curled-up fifth dimension, Klein, building on Kaluza's initial work, succeeded in unifying electromagnetism and gravity. Apparently, the scalar in the action was considered a bit of an embarrassment in the 1920's, but in recent years Kaluza-Klein theory has experienced a revival as the expanding notions of string theory have created a need for defining actions in higher dimensions. It is interesting to note that not only did Kaluza and Klein succeed in unifying electromagnetism and gravity, but matter and geometry as well, as the photon appeared in four dimensions as a manifestation of empty five-dimensional space-time.[27]

---

[26] S. Weinberg, 1972.
[27] J. M. Overduin and P.S. Wesson, 1997.

## The Fifth Dimension Revisited

In recent years, the concept of five-dimensional gravity has been revisited by a consortium of researchers led by Paul Wesson at the University of Waterloo in Ontario, Canada. The consortium also includes members of Stanford University's Gravity Probe-B program.[28] The major difference is that in the consortium's research, the fifth dimension is not compactified. The result of this is that new terms enter into the physics, even at low energies. In standard four dimensional space-time, these terms appear as matter and energy. By moving them to the right-hand side of the four dimensional equations they provide an induced energy-momentum tensor. According to Wesson, they have shown that, in fact, no five dimensional energy-momentum tensor is required. The results can include new forms of matter ultimately uniting gravity with its source, as well as with other fields.

An interesting point that Overduin and Wesson have shown is that should $\phi$ be constant and the electromagnetic potential be set to zero, $A_\mu = 0$, the result is a Brans-Dicke-type scalar field theory. The resulting metric can be written as:

$$\hat{g}_{AB} = \begin{pmatrix} g_{\alpha\beta} & 0 \\ 0 & \phi^2 \end{pmatrix} \qquad (13).$$

Combining this with the field equations and Kaluza's assumptions, the action integral becomes:

$$S = -\frac{1}{16\pi G} \int d^4 x \sqrt{-g} R \phi \qquad (14).$$

Compare this with equation 12. Neglecting the constant in front of the integral (as Derix and van der Schaar have done), and making the assumptions we have made with regard to the potential, $A$, and $\phi$, we see that equation 14 is a direct result of equation 12 (we show this to merely bridge the methods of Derix & van der Schaar and Overduin & Wesson – and we should also note that $F_{\mu\nu} \propto A_\mu$ which allows us to drop the last term in equation 12)[29].

Overduin and Wesson show, through a Kaluza-Klein ansatz metric, that for the metric to satisfy Einstein's equations in 4+$d$ dimensions, the Killing vectors must be independent of the extra coordinate, which means that the compact manifold is *flat*. Ultimately, they show that $\hat{g}_{\mu\nu}$ must also be flat.[30] So we flip-flop from "Kleinian" assumptions to Euclidean. Conventional compactification models require either that the extra dimensions be under a state of constant curvature or must include other modifications such as torsion or higher-derivative terms.  This is where Overduin and Wesson begin laying the groundwork for a non-compactified five-dimensional theory.

---

[28] See http://astro.uwaterloo.ca/~wesson for more information on the consortium.
[29] C.W. Misner, K.S. Thorne, and J.A. Wheeler, 1973.
[30] J.M. Overduin and P.S. Wesson, 1997.

The groundwork for their development is the *dependence* of physical quantities on the fifth coordinate.  This is something we have not seen as yet in our development of the fifth dimension.  This dependence is precisely what produces the electromagnetic radiation as well as a general form of matter from geometry via the higher-dimensional field equations.[31]  One of the interesting outcomes of this research is that it ultimately becomes a more easily testable theory.  The compactified dimensions of Klein are not necessarily lengthlike in nature, but the new non-compactified dimensions, with the cylindrical condition removed, can be represented as lengthlike.  Another point of interest is that previous authors have maintained Klein's mechanism of harmonic expansion which means the compact manifold must have finite volume.  With non-compactified dimensions, no such requirement exists.

Overduin and Wesson thus write the metric as:

$$(\hat{g}_{AB}) = \begin{pmatrix} g_{\alpha\beta} & 0 \\ 0 & \varepsilon\phi^2 \end{pmatrix} \qquad (15).$$

The $\varepsilon$ term is introduced to allow for a timelike as well as a spacelike signature for the fifth dimension requiring only that $\varepsilon^2 = 1$.  Please note that there is a difference between timelike and temporal here.  Having a timelike signature does not mean the dimension is necessarily temporal (and thus non-causal).  Time has rarely been considered in compactified dimensions due to a variety of problems that arise from its inclusion.  However, in non-compactified theories, some of these problems vanish.

The components of the Ricci tensor can then be represented as:

$$\hat{R}_{\alpha\beta} = R_{\alpha\beta} - \frac{\nabla_\beta(\partial_\alpha\phi)}{\phi} + \frac{\varepsilon}{2\phi^2}\left(\frac{\partial_4\phi\partial_4 g_{\alpha\beta}}{\phi} - \partial_4 g_{\alpha\beta} + g^{\gamma\delta}\partial_4 g_{\alpha\gamma}\partial_4 g_{\beta\delta} - \frac{g^{\gamma\delta}\partial_4 g_{\gamma\delta}\partial_4 g_{\alpha\beta}}{2}\right)$$

$$\hat{R}_{\alpha 4} = \frac{g^{44}g^{\beta\gamma}}{4}\left(\partial_4 g_{\beta\gamma}\partial_\alpha g_{44} - \partial_\gamma g_{44}\partial_4 g_{\alpha\beta}\right) + \frac{\partial_\beta g^{\beta\gamma}\partial_4 g_{\gamma\alpha}}{2} + \frac{g^{\beta\gamma}\partial_4(\partial_\beta g_{\gamma\alpha})}{2} - \frac{\partial_\alpha g^{\beta\gamma}\partial_4 g_{\beta\gamma}}{2}$$

$$- \frac{g^{\beta\gamma}\partial_4(\partial_\alpha g_{\beta\gamma})}{2} + \frac{g^{\beta\gamma}g^{\delta\varepsilon}\partial_4 g_{\gamma\alpha}\partial_\beta g_{\delta\varepsilon}}{4} + \frac{\partial_4 g^{\beta\gamma}\partial_\alpha g_{\beta\gamma}}{4}$$

$$\hat{R}_{44} = -\varepsilon\phi\ \phi - \frac{\partial_4 g^{\alpha\beta}\partial_4 g_{\alpha\beta}}{2} - \frac{g^{\alpha\beta}\partial_4(\partial_4 g_{\alpha\beta})}{2} + \frac{\partial_4\phi g^{\alpha\beta}\partial_4 g_{\alpha\beta}}{2\phi} - \frac{g^{\alpha\beta}g^{\gamma\delta}\partial_4 g_{\gamma\beta}\partial_4 g_{\alpha\delta}}{4} \quad (16)$$

---

[31] Ibid.

Assuming that no higher-dimensional matter exists (that's a tricky line we're going to avoid crossing), the four-dimensional Ricci tensor becomes:

$$R_{\alpha\beta} = \frac{\nabla_\beta(\partial_\alpha\phi)}{\phi} - \frac{\varepsilon}{2\phi^2}\left(\frac{\partial_4\phi\partial_4 g_{\alpha\beta}}{\phi} - \partial_4(\partial_4 g_{\alpha\beta}) + g^{\gamma\delta}\partial_4 g_{\alpha\gamma}\partial_4 g_{\beta\delta} - \frac{g^{\gamma\delta}\partial_4 g_{\gamma\delta}\partial_4 g_{\alpha\beta}}{2}\right) \quad (17)$$

Overduin and Wesson then write the second of equations 16 in the form of a conservation law:

$$\nabla_\beta P_\alpha^\beta = 0 \quad (18)$$

where they have defined a new four-tensor as:

$$P_\alpha^\beta \equiv \frac{1}{2\sqrt{\hat{g}_{44}}}\left(g^{\beta\gamma}\partial_4 g_{\gamma\alpha} - \delta_\alpha^\beta g^{\gamma\varepsilon}\partial_4 g_{\gamma\varepsilon}\right) \quad (19)$$

Finally, the third of equations 16 takes the form of a scalar wave equation for $\phi$:

$$\varepsilon\phi \quad \phi = -\frac{\partial_4 g^{\alpha\beta}\partial_4 g_{\alpha\beta}}{4} - \frac{g^{\alpha\beta}\partial_4(\partial_4 g_{\alpha\beta})}{2} + \frac{\partial_4\phi g^{\alpha\beta}\partial_4 g_{\alpha\beta}}{2\phi} \quad (20)$$

Equations 17 through 20 form the basis for the non-compactified five dimensional Kaluza-Klein theory developed by Overduin and Wesson.[32] The physical meaning of the components of these equations as well as their application to cosmology and astrophysics are discussed in depth in their paper. Their results can be compared to those derived by Derix and van der Schaar. The non-compactified equations are more complicated, and this has been an overriding motivation for compactification in the past, but the physical significance of the non-compactified equations is interesting to note (again, see Overduin and Wesson[33] as well as other reports from the consortium[34]).

## Classical String Theory

In order to more fully understand the next dimensional jump, it is necessary to digress for a moment into explaining some of the underlying methods of classical string theory, which primarily deals with, at least here, bosonic strings. To fully understand string theory, it is necessary to understand quantum field theory as modern string theory is simply a theory of quantum gravity. In addition, the mathematics of string theory can get phenomenally complex. In the interest of brevity, a few choice topics and equations will be presented in an attempt to give the flavor of string theory as a basis for moving into higher dimensional analysis.

---

[32] Ibid.
[33] Ibid.
[34] http://astro.uwaterloo.ca/~wesson.

String theory is a relatively recent phenomenon. The first hint at its existence came in 1968 when Gabriele Veneziano at CERN in Geneva needed a solution to a vexing problem he was working on at the time. Surprisingly, he found he was able to use a little-used purely mathematical tool developed by Leonard Euler nearly 200 years before called the Euler beta-function. The solution however lacked some sense of physical meaning or justification – i.e. no one was certain as to why it worked. However, two years later, the concept of strings was developed in the works of Yoichiro Nambu of the University of Chicago, Holger Nielsen of the Niels Bohr Institute, and Leonard Susskind of Stanford University and the physical meaning to Veneziano's problem was introduced. Nambu, Neilsen, and Susskind postulated that zero-dimensional point-particles were actually one-dimensional vibrating strings (thus instantaneously adding an additional dimension to the mix, at least in theory). They showed that the nuclear interactions of particles modeled as these strings were exactly described using the Euler beta-function.[35] The resonances of the vibrating strings determines the masses of the point particles we observe in nature.

Unfortunately, string theory sat dormant for over a decade due to inconsistencies and problems in some of the predictions. In 1984, John Schwarz of Cal Tech and Michael Green of Queen Mary College launched what is now known as "the first superstring revolution." During this three year period, from 1984 through 1986, more than one-thousand research papers were published on the subject.[36] Even in its initial form, superstring theory was able to unite the four forces in nature as well as matter. As a note, the "super" in superstring comes from the incorporation of supersymmetry into the theory (which has profound implications on the length scale of the actual strings as we will see in coming sections). For now, let us delve into a bit of the basics of classical string theory.

As we stated earlier, the concept of a string in its most basic form is that of a zero-dimensional point-particle magnified to such an extent that it is actually a one-dimensional vibrating string. Initially we will consider this string to be a closed loop (there are other string theories that will be discussed shortly that include non-closed loops). Just as a point particle draws out a worldline as it travels in space-time, a string sweeps out a world-sheet – one-dimension higher than a worldline. In this way, construction of a space-time diagram becomes more complicated. As such, $\hbar = c = 1$ are not *natural* units for strings (mass has the unit of inverse length). Additional introduced quantities include a new coupling constant in the form of a string tension, T, which has the units of $(length)^{-2}$ when $\hbar = c = 1$ which then introduces a characteristic length squared, $L^2$. Conversion to ordinary units defines this length as $L = \sqrt{c\hbar/\pi T}$. Being ultimately a theory of quantum gravity, this length must be on the order of the Planck length, $L_p = \sqrt{G\hbar/c^3}$, which is the only length that can be constructed from G, $\hbar$, and c.

---

[35] B. Greene, *The Elegant Universe: Superstrings, Hidden Dimensions, and the Quest for the Ultimate Theory*, Norton, 1999.
[36] Ibid.

Brian Hatfield has a detailed discussion of the units and scales of strings in Chapter 21 of his book.[37]

Strings also happen to be Wilson loops which are closed strings (of glu) of infinitesimal width. This model for strings goes back to its original development in the 1970's when it was used to describe properties in hadron physics. The strings were the gluons holding the bound quark states together to form hadrons. This led to work on developing QCD and Yang-Mills as string theories of Wilson loops.

If all point-particles are indeed strings, there would presumably be a string interaction here between the quarks and the gluons, both actually being strings. The oddly fascinating part of string theory is that when strings interact they simply produce another string which is, geometrically, though not necessarily topologically, the same as the strings that formed it. In this way we have no way of telling if a particular string was formed as a result of interactions simply by looking at it. The benefit of this is that standard perturbation theory is trivial since it really only would describe the topology of the new string. In addition, we are unable to detect exactly where a sting interaction has occurred as it will look different in two Lorentz frames.[38]

For given string theories there exists a maximum allowable space-time dimension beyond which the theory ultimately breaks down. The critical dimension is determined by the number of local supersymmetries on the string's world-sheet. If absolutely no supersymmetry is present, the critical dimension is $D = 26$. This is the maximum number of existing dimensions proposed by any string theory. It was in fashion as a possible solution for several years but, as we will see, Witten's "second superstring revolution" in 1995 may have doused that option. For 1 supersymmetry, the critical dimension is $D = 10$. This was the most widely accepted theory until Witten's revolution in 1995 and still forms the basis of $D = 11$ theory. It is also the theory that we will be focusing on it our brief glimpse at classical string theory.

The action for a relativistic particle of rest mass *m* is given as:

$$S = -m \int d\tau \sqrt{\dot{x}_\mu \dot{x}^\mu} \qquad (21)$$

or, in a form without the nasty square root:

$$S = -\frac{1}{2} \int d\tau \frac{1}{\lambda(\tau)} \left( \dot{x}^\mu \dot{x}_\mu \right) + \lambda(\tau) m^2 \qquad (22).$$

In this case $\tau$ is not necessarily the proper time but is instead a parametrization of the particle's world-line. Equation 22 is the Lagrange multiplier version of equation 21. Hatfield presents a more detailed description of the derivation of equation 22 from equation 21.[39] In comparison, the string action that corresponds to equation 21 (the point-particle action) is called the Nambu-Goto action. It is defined as:

$$S = -T \int d\sigma d\tau \sqrt{(\dot{x}^\mu \cdot x'_\mu)^2 - (\dot{x}^\mu)^2 (x'_\nu)^2} \qquad (23)$$

where the dotted coordinates are the customary $\tau$ derivatives and the primed coordinates are the $\sigma$ derivatives. Similarly, the equivalent string description for equation 22 is the Polykov action defined as:

$$S = -\frac{T}{2} \int d\sigma d\tau \sqrt{-g}\, g^{ab} \partial_a x^\mu \partial_b x_\mu \qquad (24).$$

---

[37] B. Hatfield, *Quantum Field Theory of Point Particles and Strings*, Addison-Wesley, 1992.
[38] Ibid.

Compare the results of the action integrals of these topological defects to equations 12 and 14, the action integrals for the five-dimensional systems of Kaluza-Klein theory. (Keep in mind the actions here are describing paths of particles and strings while the Kaluza-Klein actions describe a field). A cursory comparison shows that one of the major introductions to the string action is the string tension, *T*. However, we should note that by once again expanding from a zero-dimensional point-particle up one dimension to a one-dimensional string we get an equation (24) that is suspiciously similar to equation 14, the Overduin and Wesson action integral.

## Supergravity and Superstrings: 10 or 11 dimensions?

In expanding beyond five dimensions, we actually explode into more than double that. This is based on the critical dimension we mentioned in the previous section. String theory got "stuck in the mud" for many years, in particular in the early 1990's, within a maze of infinities and other odd problems including the lack of a consensus on the value of the critical dimension. In 1995, however, Edward Witten of the Institute for Advanced Study in Princeton, New Jersey, presented a seminal lecture at the Strings 1995 conference at USC that launched the "second superstring revolution." The meat of his lecture showed that a minimum of 11 dimensions was required for a Kaluza-Klein theory to unify all of the forces in the standard model of particle physics (namely it contained the gauge groups of the strong and electroweak interactions). Prior to this, 11 was precisely the same number of dimensions determined by Nahm to be a *maximum* for consistency with, none other than, the graviton (with a maximum of spin 2)![40] So, it seems, the unification of the four forces of nature necessarily *required* 11 dimensions! In fact, there were even more conditions that were discovered to apply that fixed the critical dimension at 11. In addition, the four dimensions of the visible world split out perfectly from the total 11 leaving 7 compactified or non-physical dimensions in its wake.

The supergravity concept was developed to add the extra matter fields to the equations. The easiest way to do this was to make the theory supersymmetric which means every boson has some, as yet undiscovered, fermionic superpartner. This very development poses an enormous barrier to the experimental verification of string theory, however. Obtaining the energy levels necessary to produce these supermassive superpartners in a laboratory is well beyond our current reach. This isn't the only problem with the $D = 11$ supergravity theory. One major problem that turned up is that the compact manifolds did not produce quarks and leptons. Several other issues involving chirality and a rather large cosmological constant also arise.

---

[39] Ibid.
[40] J.M. Overduin and P.S. Wesson, 1997.

The real breakthrough came with the development of two separate ten-dimensional supergravity models that were able to solve the anomaly problems while also maintaining the uniqueness that the eleven dimensional theory held (the minimum/maximum critical dimension issues). These two theories were based on the groups $SO(32)$ and $E_8 \times E_8$. The extra terms that needed to be added corresponded to those that appeared naturally in low-energy superstring theory. The first sign of trouble with these two theories is that they predicted five separate string theories between them. However, Witten has proposed an entirely new theory called M-theory (M for membrane) that unites the five complete string theories along with supergravity under one umbrella: Type I (the bosonic string theory we looked at in the previous section), Type IIA, Type IIB, Heterotic-O ($O_8 \times O_8$), Heterotic-E ($E_8 \times E_8$), and $D = 11$ supergravity. Details of this unification depend on the introduction of a new concept into the fray: that of duality.

### Duality and M-Theory

Duality was really the essence of Witten's lecture at Strings 1995. The idea behind duality is that a singular physical system can be described by two seemingly separate theories. More to the point, it's like looking at a house from the front and then from the back. Initially there might be no indication that you're actually looking at one-in-the-same house when, in fact, further research eventually proves it is indeed one house. One fantastically interesting application of duality in physics is that when shrinking down to the scale of the Planck length while looking at a circular dimension of radius $R$, once we pass through the Planck length we find that the physics described by the system with radius $1/R$ is precisely the same as that described by the radius $R$. Therefore, essentially, the universe at sub-Planck scales on the order of say something as absurd as $10^{-80}$ m is exactly the same as the universe at $1/10^{-80}$ m (which is huge). So we see that the Planck length mirrors us back outward if we try to continue to probe to smaller lengths, all thanks to the introduction of duality. This means that there is an exact lower limit to the size of compactified dimensions – the Planck length (radial in this example).

Another interesting artifact of duality is the fact that the exact shape of the compactified dimensions (taking the form of a Calabi-Yau space as we will see) is not necessarily important. Two completely different shapes can produce the exact same physics. Witten used this idea to develop M-Theory, proposing that the different superstring theories as well as $D = 11$ supergravity were all portions of the same theory that appeared different simply on the surface but, thanks to duality, described exactly the same physics. To couple the various theories to each other and to M-Theory as a whole, dualities have been employed to show that Type I and Heterotic-O are coupled, while Heterotic-O is also coupled to Heterotic-E which is coupled to M-Theory's core, which is coupled to Type-IIA which is coupled to Type-IIB which is finally coupled to itself. More work is being performed in this area in an effort to unite supergravity and also to more clearly develop the exact form of M-Theory. Of particular interest to us in regard to this paper is what all this has to say about the extra dimensions it offers us.

### Traversable Dimensions and F-Theory

In superstring theory the extra 7 dimensions are compactified into a complex set of shapes that is dictated by the equations of the theory. It turns out the geometrical shapes dictated by string theory satisfied a previously known set of geometrical spaces known as Calabi-Yau spaces (after Eugenio Calabi of the University of Pennsylvania and Shing-Tung Yau of Harvard University). The mathematics of Calabi-Yau shapes is quite complex and a visual representation of 7 spatial dimensions on a sheet of paper is quite complicated (though, see Greene[41] page 207 for a reasonable approximation) so we will not delve deeper into them here. As we stated in the previous section, duality allows for a veritable zoo of Calabi-Yau shapes that ultimately describe the exact same physics.

The physics described by Calabi-Yau spaces in string theory is actually indirectly experimentally testable. As we stated earlier, the resonances of the vibrating strings determines the masses of the elementary particles in physics. The strings are free to vibrate in virtually any direction in the spatially extended dimensions and can also vibrate within the compactified dimensions. However, when vibrating in the compactified dimensions, the precise nature of the Calabi-Yau space describing the higher dimensions constrains the motion of the vibrating string. So in understanding the precise Calabi-Yau spatial geometry of a particular manifold, additional constraints can be placed on the strings making it theoretically easier to determine the precise physical nature of the string – e.g. the mass and charge of the particle it describes. Physicists consider this to be one of the most far-reaching and profoundly insightful facts of string theory. Additional work by Andrew Strominger and others allowed for the slight modification of this theory to solve the problem of collapsing dimensions. In this theory, a one-dimensional string is called a one-brane and can completely surround a one-dimensional piece of space. If this one-dimensional string is blown up like an inner-tube or a tire it becomes two-dimensional and is called a two-brane. A two-brane can completely surround a two-dimensional piece of space. One can easily see where this is heading. The idea is that by surrounding the extra spatial dimensions with a brane (a multi-dimensional string) the cataclysmic effects of collapse can be blocked.[42]

Based on these concepts, the compactified dimensions are traversable by strings, but not by anything larger. So technically to us the extra dimensions are not traversable. Objects on the order of a point-particle (as we see them) and larger can only traverse the four non-compactified dimensions. The nature of these dimensions is not as well known as we think. The precise nature of the Euclidean spatial dimensions appears locally to be flat, though general relativity has shown that the manifold of these three dimensions along with time can be bent and warped in the presence of gravity. On a larger scale the precise shape of the universe has a profound effect on the ultimate shape of the Euclidean dimensions. Technically speaking the most interesting case would be a closed universe in which the Euclidean dimensions eventually bent back on themselves (legitimate physicists have recently suggested this could be possible[43]). Unfortunately for those who find this notion romantic, a recent paper by P. De Bernardis, et. al. in the journal *Nature* based on balloon research in Antarctica has shown that the universe is indeed flat (Euclidean - at least for now).

---

[41] B. Greene, 1999.
[42] Ibid.

The fifth dimension as described by the original Kaluza-Klein theory would not be traversable except possibly by vibrating strings. However, the new non-compactified fifth dimension as proposed by Wesson's consortium might allow for a fifth fully traversable dimension. Whether this dimension is truly Euclidean as well based on the recent observations of De Bernardis, et. al. would need to be probed.

Standard traversable Euclidean dimensions have two degrees of physical freedom. To our knowledge, temporal dimensions do not. Based on standard causal-based physics, the only time dimension that we are aware of has a single degree of freedom – forward. Science fiction writers and some physicists have speculated that time travel is possible. An extensive base of scientific research has been performed on wormholes, some of which suggests the possibility of time travel, though the research has taken this to be a useful mathematical tool rather than speculating on its actual physical existence (see Visser for a detailed look at current wormhole research and the mathematical foundations for theoretical time travel[44]). M.J. Duff of the University of Michigan began his compilation on string theory with the following quote from Mother Goose: "Nature requires five, Custom allows seven, Idleness takes nine, And wickedness eleven." The appropriateness of this quote became apparent when Cumrun Vafa of Harvard in February of 1996 first suggested F-Theory (building on work by a number of others – for a very interesting and detailed overview of F-Theory, David R. Morrison at Duke, a leading string theorist and F-Theorist, has archived six lectures on RealVideo on his website[45]). In this theory, the 11 dimensions of M-Theory are extended to 12 in order to solve a few select problems inherent in M-Theory. The interesting thing is that this additional dimension is *temporal*. Immediately, the mere philosophical implications are staggering if there is physical fact lying behind the mathematics. But, for now, let's remain with the idea that there is a single degree of freedom in all possible temporal dimensions combined and simply say that F-Theory is a convenient mathematical tool.

## Creation and Conclusion

One final point in discussing this dizzying array of dimensions is to briefly mention how they formed. At some point a few fractions of a second after the Big Bang, spontaneous symmetry breaking occurred causing the non-compactified dimensions to expand while the compactified ones curled up into a Calabi-Yau "ball." This was a result of the presence of a tremendous amount of tension that, when released during the symmetry breaking, caused the dimensions to "snap into position." Whether these dimensions will be united at some time in the tremendously distant future is of course unknown. But what we have learned over the last two millennia (a short time, in perspective) is tremendous. We have slowly developed, dimension by dimension, a world of multiple dimensions, some seen, some unseen. The implications of the physics and topology of these dimensions are far reaching and years of research are still ahead of us. Ultimately, the reward should be well worth the hunt but "only time will tell…"

---

[43] Ibid.
[44] M. Visser, 1996.
[45] See http://www.cgtp.duke.edu/~drm/ftheory/ to access the RealVideo lectures.

# Bibliography

**Print Resources - Technical**